\documentclass{article}

\usepackage{amsmath,amssymb,amsthm}

\newtheorem{theorem}{Theorem}[section]
\newtheorem{corollary}[theorem]{Corollary}

\newtheorem{lemma}[theorem]{Lemma}

\theoremstyle{remark}
\newtheorem{remark}{Remark}[section]

\theoremstyle{definition}
\newtheorem{definition}{Definition}[section]

\begin{document}

\title{Necessary Optimality Conditions for Fractional
Action-Like Integrals of Variational Calculus with
Riemann-Liouville Derivatives of Order
$\left(\alpha,\beta\right)$\footnote{This is a preprint of an
article accepted for publication (28-02-2007) in \emph{Math.
Methods Appl. Sci.}, Wiley,
\texttt{http://www.interscience.wiley.com}}}

\author{Rami Ahmad El-Nabulsi${}^{\dag}$\\
        \texttt{atomicnuclearengineering@yahoo.com}
        \and
        Delfim F. M. Torres${}^{\ddag}$\\
        \texttt{delfim@mat.ua.pt}}

\date{${}^{\dag}$Department of Nuclear and Energy Engineering\\
      Faculty of  Mechanical, Energy and Production Engineering\\
      Cheju National University, 690-756 Jeju, South Korea\\[0.3cm]
      ${}^{\ddag}$Centre for Research on Optimization and Control\\
      Department of Mathematics, University of Aveiro\\
      Campus Universit\'{a}rio de Santiago, 3810-193 Aveiro, Portugal}

\maketitle


\begin{abstract}
We derive Euler-Lagrange type equations for fractional action-like
integrals of the calculus of variations which depend on the
Riemann-Liouville derivatives of order
$\left(\alpha,\beta\right)$, $\alpha > 0$, $\beta > 0$, recently
introduced by J.~Cresson and S.~Darses. Some interesting
consequences are obtained and discussed.
\end{abstract}

\smallskip


\smallskip


\smallskip



\section{Introduction}

Fractional derivatives and integrals play a leading role in the
understanding of complex, classical or quantum, conservative or
nonconservative, dynamical systems with holonomic as well as with
nonholonomic constraints. Many applications of fractional
differentiation and integration can be found in turbulence and
fluid dynamics, chaotic dynamics, solid state physics, chemistry,
stochastic dynamical systems, plasma physics and controlled
thermonuclear fusion, kinetic theory, quantization, field theory,
nonlinear control theory, image processing, nonlinear biological
systems, material sciences, rheological properties of rocks,
scaling phenomena, astrophysics, etc. (see \textrm{e.g.}
\cite{[41],[43],[26]}). Its origin goes back more than three
centuries, when in 1695 L'Hopital made some remarks to Leibniz
about the mathematical meaning of a fractional derivative of order
$1/2$. Leibniz's response was ``\emph{an apparent paradox, from
which one day useful consequences will be drawn}''. In these words
fractional calculus was born. It was primarily a study reserved to
the best minds in mathematics, including J.~Fourier, N.~H.~Abel,
J.~Liouville, B.~Riemann, H.~Holmgren, etc., who contributed
strongly to the fractional analysis program. A deep research was
carried out by Liouville from 1832 to 1837, where he succeeded in
defining the first fractional integration operator. Later on,
further developments lead to the construction of the well-known
Riemann-Liouville fractional integral operator, which plays
nowadays a leading and important role in the analysis of complex
dynamical systems. There exist today many different forms of
fractional integral operators, ranging from divided-difference
types to infinite-sum types, including the Grunwald-Letnikov
fractional derivative, the Caputo fractional derivative, etc. One
can say, however, that the Riemann-Liouville operator is still one
of the most frequently used for fractional integration.

The fractional calculus was considered a theoretical mathematical
field with no physical applications for more than three centuries.
Last decades have shown the contrary. Several fields of
application of fractional differentiation and fractional
integration are already well established, some others have just
started. One of the areas that recently emerged within the
fractional framework and which is being subject to strong research
is the Calculus of Variations (CoV) and respective Euler-Lagrange
type equations. In 1996 F.~Riewe formulated the CoV problem with
fractional derivatives and obtained the respective Euler-Lagrange
equations (ELe's), combining both conservative and
non-conservative cases \cite{[33]}. In 2001 another approach was
developed by M.~Klimek by considering fractional problems of the
CoV but with symmetric fractional derivatives. Correspondent ELe's
were obtained, using both Lagrangian and Hamiltonian formalisms
\cite{[35]}. In 2002 O.~Agrawal extended Klimek variational
problems by considering right and left fractional derivatives in
the Riemann-Liouville sense \cite{[36]}. In 2004 Agrawal ELe's
were used by D.~Baleanu and T.~Avkar to investigate problems with
Lagrangians which are linear in the velocities \cite{[29]}. In all
the above mentioned studies, ELe's depend on left and right
fractional derivatives, even when the problem depends only on one
type of them. To avoid this situation, in 2005 M.~Klimek studied
problems depending on symmetric derivatives, proving that in this
case ELe's include only the derivatives that appear in the
formulation of the problem \cite{[30]}. One major problem with all
the above mentioned approaches is the presence of a non-local
fractional differential operator with an adjoint which is not its
symmetric. Other difficulties arise due to a complicated Leibniz
rule and the non-existence of a fractional analogue to the chain
rule.

The physical reasons for the appearance of fractional equations
are, in general, long-range dissipation and non-conservatism. For
this reason, it seems of interest to study the fractional
Hamiltonian of nonconservative dynamical systems. With this in
mind, the first author has proposed in 2005 a novel approach,
entitled \emph{Fractional Action-Like Variational Approach}
(FALVA), to model nonconservative dynamical systems
\cite{[44],[45]}. The Euler-Lagrange equations proved to be
similar to the classical ones, with no fractional derivatives
appearing, but with the presence of a fractional generalized
external force acting on the system. The momentum, the Hamiltonian
and Hamilton's equations were proved to depend on the fractional
order of integration and to vary inversely to time. In the present
work we extend FALVA formalism and previous results by considering
Riemann-Liouville fractional derivatives of order
$\left(\alpha,\beta\right)$. Respective Euler-Lagrange equations
are obtained using a Taylor-Riemann series expansion; the concept
of fractional constant of motion is introduced and illustrated.


\section{Fractional Euler-Lagrange Equations}

We begin by recalling the standard definitions of
Riemann-Liouville fractional integrals and fractional derivatives.

\begin{definition}[Left Riemann-Liouville fractional integral]
If $f\left(t\right) \in C\left({a ,b}\right)$, and $\alpha > 0$,
then the left Riemann-Liouville fractional integral of order
$\alpha$ is defined and denoted by
\begin{equation*}
I_{a_+}^\alpha  f\left( t \right) = \frac{1}{{\Gamma \left( \alpha
\right)}}\int\limits_{a}^t {f\left( \tau \right)\left( {t - \tau }
\right)^{\alpha  - 1} d\tau } \, .
\end{equation*}
\end{definition}

\begin{definition}[Right Riemann-Liouville fractional integral]
If $f\left(t\right) \in C\left(a,b\right)$, and $\alpha > 0$, then
the right Riemann-Liouville fractional integral of order $\alpha$
is defined and denoted by
\begin{equation*}
I_{b_-}^\alpha  f\left( t \right) = \frac{1}{{\Gamma \left( \alpha
\right)}}\int\limits_{t}^b {f\left( \tau \right)\left( {\tau - t }
\right)^{\alpha  - 1} d\tau } \, .
\end{equation*}
\end{definition}

\begin{definition}[Left and Right Riemann-Liouville fractional derivatives]
If $f\left( t \right) \in C^1 \left( {a ,b} \right)$, and $\alpha
> 0$, then the left and right Riemann-Liouville fractional
derivatives of order $\alpha$, denoted respectively by $D_{a_ +
}^\alpha$ and $D_{b_- }^\alpha$, are defined by
\begin{gather*}
D_{a_ +  }^\alpha  f\left( t \right) = \frac{1}{{\Gamma \left( {n
- \alpha } \right)}}\left(\frac{d}{{dt}}\right)^n
\int\limits_{a}^t {f\left(
\tau \right)\left( {t - \tau } \right)^{ n - \alpha - 1} d\tau } \, , \\
D_{b_ -  }^\alpha  f\left( t \right) = \frac{1}{{\Gamma \left( {n
- \alpha } \right)}}\left(-\frac{d}{{dt}}\right)^n
\int\limits_{t}^b {f\left( \tau \right)\left( {t - \tau }
\right)^{n - \alpha -1} d\tau } \, ,
\end{gather*}
where $n$ is such that $n-1 \le \alpha < n$.
\end{definition}

The classical integrals, and derivatives of order one, are
obtained setting $\alpha = 1$. For an introduction to the
fractional calculus we refer the reader to \cite{[10],[12],[8]}.

In this work we consider the fractional derivative operator of order
$\left(\alpha,\beta\right)$ recently introduced by J.~Cresson and S.~Darses
\cite{CressonDarses} (see also \cite{[37]}):

\begin{definition}
Given $a$, $b \in \mathbb{R}$, $a < b$, and $\gamma  \in
\mathbb{C}$, the fractional derivative operator of order
$\left(\alpha,\beta\right)$, $\alpha,\, \beta  > 0$, is defined by
\begin{equation*}
D_\gamma^{\alpha ,\beta} = \frac{1}{2}\left[ {D_{a_ +  }^\alpha -
D_{b_ -  }^\beta  } \right] + \frac{{i\gamma }}{2}\left[ {D_{a_ +
}^\alpha   + D_{b_ -  }^\beta  } \right]
\end{equation*}
where $i = \sqrt{- 1}$.
\end{definition}

\begin{remark}
The operator $D_\gamma^{\alpha ,\beta}$ extends the classical
Riemann-Liouville fractional derivatives: for $\gamma = -i$ we
have $D_\gamma^{\alpha ,\beta} = D_{a_ +  }^\alpha$; for $\gamma =
i$ we obtain $D_\gamma^{\alpha ,\beta} = - D_{b_ -  }^\beta$.
\end{remark}

The following Lemma is useful: we make use of \eqref{eq:6}
to prove the fractional Euler-Lagrange equations
associated to our problem (\textrm{cf.} proof of Theorem~\ref{thm1}).

\begin{lemma}[\cite{[37]}]
\label{lem1} If $f,\, g \in C^1$ with $f(a) = f(b) = 0$ or $g(a) =
g(b) = 0$, then
\begin{equation}
\label{eq:6}
\int\limits_a^b {D_\gamma ^{\alpha ,\beta } } f\left( t \right)g\left( t \right)dt
=  - \int\limits_a^b {f\left( t \right)D_{ - \gamma }^{\beta ,\alpha } } g\left( t \right)dt \, .
\end{equation}
\end{lemma}

We are now in conditions to formulate the
$\left(\alpha,\beta\right)$ fractional action-like
variational problem.

\begin{definition}
\label{def:Prb}
Consider a smooth manifold $M$ and let $L$ be
a smooth Lagrangian function
$L: \mathbb{C}^d \times \mathbb{R}^d \times \mathbb{R} \to \mathbb{R}$,
$d \ge 1$. For any piecewise smooth path
$q:\left[ {a, b } \right] \to M$ satisfying fixed boundary
conditions $q(a) = q_a$ and $q(b) = q_b$ we define the following
fractional action integral:
\begin{equation}
\label{eq:7} S_{\gamma ,\left( {a,b} \right)}^{\alpha ,\beta
}\left[ q \right] = \frac{1}{{\Gamma \left( \alpha
\right)}}\int\limits_a^b {L\left( {D_\gamma ^{\alpha ,\beta }
q\left( \tau  \right), q\left( \tau  \right),\tau } \right)\left(
{t - \tau } \right)^{\alpha  - 1} d\tau } \, .
\end{equation}
The $\left(\alpha,\beta\right)$ fractional action-like variational problem
consists in finding an admissible $q(\cdot)$ which minimizes \eqref{eq:7}.
\end{definition}

\begin{remark}
We consider two time variables: the intrinsic time $\tau$ and the
observer time $t$. This multi-time characteristic is important in
applications and is the main ingredient of the theory being developed
by C.~Udriste \cite{Udriste}.
\end{remark}

\begin{remark}
The fractional action integral \eqref{eq:7} is a generalization
of the FALVA action integral of \cite{[44],[45],[46]}.
\end{remark}

Theorem~\ref{thm1} gives a necessary optimality condition for
$q(\cdot)$ to be a solution of the $\left(\alpha,\beta\right)$
fractional action-like variational problem.

\begin{theorem}[$\left(\alpha,\beta\right)$ fractional Euler-Lagrange equations]
\label{thm1}
If $q:\left[ {a, b } \right] \to M$ is a minimizer
of the $\left(\alpha,\beta\right)$ fractional
action-like variational problem (\textrm{cf.} Definition~\ref{def:Prb}), then
\begin{multline}
\label{eq:8}
\frac{{\partial L}}{{\partial q}}\left( {D_\gamma ^{\alpha ,\beta }
q\left( \tau  \right),q\left( \tau  \right),\tau } \right)
- D_{ - \gamma ;\tau }^{\beta ,\alpha } {\frac{{\partial L}}{{\partial
\dot q}}}\left( {D_\gamma ^{\alpha ,\beta }
q\left( \tau  \right),q\left( \tau  \right),\tau }\right) \\
= \frac{{1 - \alpha }}{{t - \tau }} \, \frac{{\partial L}}{{\partial
\dot q}}\left( {D_\gamma ^{\alpha ,\beta } q\left( \tau  \right),
q\left( \tau  \right),\tau } \right)
\end{multline}
where $D_{ - \gamma ;\tau }^{\beta ,\alpha }$
represents the fractional derivative with respect to time $\tau$.
\end{theorem}

\begin{remark}
Theorem~\ref{thm1} is an extension of the Euler-Lagrange equations
derived in FALVA: one just needs to choose $\beta  = 1$ in \eqref{eq:8}
to obtain the Euler-Lagrange equations of \cite{[44],[45]}.
\end{remark}

\begin{proof}
We perform a small perturbation of the generalized coordinates as
$q \to q + \varepsilon h$, $\varepsilon  \ll 1$. As a result,
$D_\gamma ^{\alpha ,\beta } \left( {q + \varepsilon h} \right)
= D_\gamma ^{\alpha ,\beta } q + \varepsilon D_\gamma ^{\alpha ,\beta } h$ and
\begin{equation*}
S_{\gamma ,\left( {a,b} \right)}^{\alpha ,\beta }\left[ {q
+ \varepsilon h} \right] = \frac{1}{{\Gamma \left( \alpha  \right)}}
\int\limits_a^b {L\left( {D_\gamma ^{\alpha ,\beta } q
+ \varepsilon D_\gamma ^{\alpha ,\beta } h,q
+ \varepsilon h,\tau } \right)\left( {t - \tau } \right)^{\alpha  - 1} d\tau }
\end{equation*}
which, doing a Taylor expansion of $L\left( {D_\gamma ^{\alpha
,\beta } q + \varepsilon D_\gamma ^{\alpha ,\beta } h,q +
\varepsilon h,\tau } \right)$ in $\varepsilon$ around zero, and
integrating by parts, imply that
\begin{multline*}
S_{\gamma ,\left( {a,b} \right)}^{\alpha ,\beta } \left[ {q
+ \varepsilon h} \right] = S_{\gamma ,\left( {a,b} \right)}^{\alpha ,\beta } \left[ q \right] \\
- \frac{\varepsilon }{{\Gamma \left( \alpha  \right)}}\int\limits_a^b {\left[ {
- \frac{{\partial L}}{{\partial q}}\left( {D_\gamma ^{\alpha ,\beta }
q\left( \tau  \right),q\left( \tau  \right),\tau } \right) \, \left( {t
- \tau } \right)^{\alpha  - 1} h\left( \tau  \right)} \right.} \\
+ \left( {t - \tau } \right)^{\alpha  - 1} \frac{{\partial L}}{{\partial
\dot q}}\left( {D_\gamma ^{\alpha ,\beta }
q\left( \tau  \right)D_\gamma ^{\alpha ,\beta } h\left( \tau  \right),
q\left( \tau  \right),\tau } \right) D_{\gamma ;\tau }^{\alpha,
\beta } h\left( \tau  \right) \\
\left. { + \left( {t - \tau } \right)^{\alpha  - 1} \frac{{\partial L}}{{\partial
\dot q}}\left( {D_\gamma ^{\alpha,
\beta } q\left( \tau  \right)D_\gamma^{\alpha,\beta } h\left( \tau  \right),
q\left( \tau  \right),\tau } \right) D_\gamma ^{\alpha,
\beta } h\left( \tau  \right)} \right]d\tau  + O\left( \varepsilon  \right) \, .
\end{multline*}
Making use of Lemma~\ref{lem1} and the least action principle
we arrive to \eqref{eq:8}.
\end{proof}

\begin{definition}
A path $q:\left[ {a, b } \right] \to M$ satisfying equation \eqref{eq:8}
is said to be a \emph{fractional extremal} associated to the Lagrangian $L$.
\end{definition}

\begin{definition}
\label{def:dec:fric:f}
The right hand side of \eqref{eq:8},
\begin{equation}
\label{eq:9}
F_{\gamma ,\tau }^{\alpha ,\beta }
=  {\frac{{\alpha  - 1}}{T}} \,
\frac{{\partial L}}{{\partial
\dot q}}\left( {D_\gamma ^{\alpha ,\beta } q\left( \tau  \right),
q\left( \tau  \right),\tau } \right) \, ,
\end{equation}
$T = \tau  - t$, defines the \emph{fractional decaying friction force}.
\end{definition}

\begin{remark}
The fractional decaying friction force
satisfy the asymptotic property
$\mathop {\lim }\limits_{T \to \infty } F_{\gamma ,\tau }^{\alpha ,\beta }  = 0$.
\end{remark}


\section{The Fractional Hamiltonian Formalism}

We now consider a more general class of fractional optimal control problems:
\begin{equation}
\label{eq:15}
\begin{gathered}
S_{\gamma ,\left( {a,b} \right)}^{\alpha ,\beta } \left[ {q,u} \right]
= \frac{1}{{\Gamma \left( \alpha  \right)}}\int\limits_a^b
{L\left( {u\left( \tau  \right),q\left( \tau
\right),\tau } \right)\left( {t - \tau } \right)^{\alpha  - 1} d\tau } \longrightarrow \min \, ,\\
D_\gamma ^{\alpha ,\beta } q\left( \tau  \right)
= \varphi \left( {u\left( \tau  \right),q\left( \tau  \right),\tau } \right) \, ,
\end{gathered}
\end{equation}
$a$, $b \in \mathbb{R}$, $a < b$. In the particular case where
$\varphi \left( {u,q,\tau } \right) = u$ \eqref{eq:15} reduces to \eqref{eq:7}.
To obtain a necessary optimality condition to problem \eqref{eq:15}
we introduce the augmented action integral (\textrm{cf. e.g.} \cite{[2]}):
\begin{multline}
\label{eq:16}
S_{\gamma ,\left( {a,b} \right)}^{\alpha ,\beta }
\left[ {q,u,p^{\alpha,\beta}  } \right] =
\frac{1}{{\Gamma \left( \alpha  \right)}}\int\limits_a^b
\Bigl[ \mathcal{H}^{\alpha,\beta}  \left( {u\left( \tau  \right),
q\left( \tau  \right),p^{\alpha,\beta}  \left( \tau  \right),\tau } \right)\\
- p^{\alpha,\beta}  \left( \tau  \right)D_\gamma ^{\alpha ,\beta }
q\left( \tau  \right) \Bigr]d\tau
\end{multline}
where $p^{\alpha,\beta}$ is the fractional Lagrange multiplier
and the fractional Hamiltonian $\mathcal{H}^{\alpha,\beta}$ is defined by
\begin{equation*}
\mathcal{H}^{\alpha,\beta}\left( {u,q,p^{\alpha,\beta}  ,\tau } \right)
= L\left( {u,q,\tau } \right)\left( {t - \tau } \right)^{\alpha  - 1}
+ p^{\alpha,\beta}  \varphi \left( {u,q,\tau } \right) \, .
\end{equation*}

\begin{theorem}
\label{th3}
If $\left( {q,u} \right)$  is a minimizer of problem \eqref{eq:15},
then there exists a co-vector function $p^{\alpha,\beta}$
such that the following conditions hold:
\begin{itemize}
\item the fractional Hamiltonian system
\begin{equation}
\label{eq:17}
\left\{ \begin{array}{l}
 D_\gamma ^{\alpha ,\beta } q\left( \tau  \right)
 = \frac{{\partial \mathcal{H}^{\alpha,\beta}}}{{\partial p^{\alpha,\beta}  }}\left(
 {u\left( \tau  \right),q\left( \tau  \right),
 p^{\alpha,\beta}  \left( \tau  \right),\tau } \right) \, ,\\
 D_{ - \gamma ;\tau }^{\beta ,\alpha }
 p^{\alpha,\beta}  \left( \tau  \right)
 =  - \frac{{\partial \mathcal{H}^{\alpha,\beta}}}{{\partial q}}\left(
 {u\left( \tau  \right),q\left( \tau  \right),
 p^{\alpha,\beta}  \left( \tau  \right),\tau } \right) \, ;\\
 \end{array} \right.
\end{equation}
\item the fractional stationary condition
\begin{equation}
\label{eq:18}
\frac{{\partial \mathcal{H}^{\alpha,\beta}}}{{\partial u}}\left( {u\left(
\tau  \right),q\left( \tau  \right),
p^{\alpha,\beta}  \left( \tau  \right),\tau } \right) = 0 \, .
\end{equation}
\end{itemize}
\end{theorem}

\begin{remark}
For the fractional problem of the calculus
of variations \eqref{eq:7} one has
\begin{equation*}
\mathcal{H}^{\alpha,\beta}\left({u,q,p^{\alpha,\beta}  ,\tau } \right)
= L\left( {u,q,\tau } \right)\left( {t - \tau } \right)^{\alpha  - 1}
+ p^{\alpha,\beta}  u \, .
\end{equation*}
The fractional stationary condition \eqref{eq:18} reduces to
\begin{equation}
\label{eq:19}
p^{\alpha,\beta}  \left( \tau  \right) =
- \frac{{\partial L}}{{\partial u}}\left( {u\left( \tau  \right),
q\left( \tau  \right),\tau } \right)\left( {t - \tau } \right)^{\alpha  - 1} \, ;
\end{equation}
the first equation in \eqref{eq:17} to
\begin{equation}
\label{eq:20}
u\left( \tau  \right) = D_\gamma ^{\alpha ,\beta } q\left( \tau  \right) \, ;
\end{equation}
while the second equation in \eqref{eq:17} takes the form
\begin{equation}
\label{eq:21}
D_{ - \gamma ;\tau }^{\beta ,\alpha } p^{\alpha,\beta} \left( \tau  \right)
=  - \frac{{\partial L}}{{\partial q}}\left( {u\left( \tau  \right),
q\left( \tau  \right),\tau } \right)\left( {t - \tau } \right)^{\alpha  - 1} \, .
\end{equation}
Using \eqref{eq:19} and \eqref{eq:20} in \eqref{eq:21} we arrive to:
\begin{multline}
\label{eq:22}
D_{ - \gamma ;\tau }^{\beta ,\alpha } \left[ {\frac{{\partial L}}{{\partial u}}\left(
{D_\gamma ^{\alpha ,\beta } q\left( \tau  \right),
q\left( \tau  \right),\tau } \right)\left( {t - \tau } \right)^{\alpha  - 1} } \right]\\
= \frac{{\partial L}}{{\partial q}}\left( {D_\gamma ^{\alpha ,\beta }
q\left( \tau  \right),q\left( \tau  \right),\tau }
\right)\left( {t - \tau } \right)^{\alpha  - 1} \, .
\end{multline}
Simple calculations show that \eqref{eq:22} is equivalent to \eqref{eq:8},
that is, Theorem~\ref{th3} is a generalization of Theorem~\ref{thm1}
to the fractional optimal control problem \eqref{eq:15}.
\end{remark}

\begin{remark}
Let us define the Poisson bracket of two dynamical
quantities $f$ and $g$, with respect to coordinates $q$
and fractional-momenta $p^{\alpha,\beta}$, by
\begin{equation*}
\left\{ {f,g} \right\} =  {\frac{{\partial f}}{{\partial p^{\alpha,\beta} }}
\cdot \frac{{\partial g}}{{\partial q}}
- \frac{{\partial f}}{{\partial q}} \cdot \frac{{\partial g}}{{\partial p^{\alpha,\beta}}}}  \, .
\end{equation*}
The fractional Hamiltonian system \eqref{eq:17} can be written in the following form:
\begin{equation*}
\left\{ \begin{array}{l}
 D_\gamma ^{\alpha ,\beta } q\left( \tau  \right)
 =  \left\{ {\mathcal{H}^{\alpha,\beta},q} \right\} \, ,\\
 D_{ - \gamma ;\tau }^{\beta ,\alpha }
 p^{\alpha,\beta}  \left( \tau  \right)
 = \left\{ {\mathcal{H}^{\alpha,\beta},p^{\alpha,\beta}} \right\}  \, .\\
 \end{array} \right.
\end{equation*}
\end{remark}

\begin{proof}
Theorem~\ref{th3} is proved applying the
$\left(\alpha,\beta\right)$ fractional Euler-Lagrange equations to
the augmented action integral (\ref{eq:16}), \textrm{i.e.}
applying (\ref{eq:8}) to
\begin{equation*}
S_{\gamma ,\left( {a,b} \right)}^{\alpha
,\beta }\left[ {q,u,p^{\alpha,\beta}  } \right] =
\frac{1}{{\Gamma \left( \alpha  \right)}}\int\limits_a^b {\left[
{\frac{{\mathcal{H}^{\alpha,\beta} - p^{\alpha,\beta}  \left( \tau
\right)D_\gamma ^{\alpha ,\beta } q\left( \tau  \right)}}{{\left(
{t - \tau } \right)^{\alpha - 1} }}} \right]\left( {t - \tau }
\right)^{\alpha  - 1} d\tau } \, ,
\end{equation*}
where $\mathcal{H}^{\alpha,\beta} =
\mathcal{H}^{\alpha,\beta}\left( {u\left( \tau \right),
q\left(\tau\right),p^{\alpha,\beta}\left( \tau\right),\tau}\right)$.
The Euler-Lagrange equation with respect to $q$ gives the second
equation of the fractional Hamiltonian system \eqref{eq:17};
the Euler-Lagrange equation with respect to $u$ gives the
fractional stationary condition \eqref{eq:18}; and, finally,
the Euler-Lagrange equation with respect to $p^{\alpha,\beta}$
gives the first equation of \eqref{eq:17}.
\end{proof}

In classical mechanics, constants of motion are derived from the
first integrals of the Euler-Lagrange equations. In the
fractional case it is necessary to change the definition
of constant of motion in a proper way \cite{[2],IJAM,CD:FredericoTorres:2007}.
Here, in order to account the presence of the fractional
decaying friction force $F_{\gamma ,\tau }^{\alpha ,\beta }$ \eqref{eq:9},
we propose the following definition of fractional constant of motion.

\begin{definition}
We say that a function $C$ of $\tau$ is a \emph{fractional
constant of motion} if and only if
$D_{ - \gamma ;\tau }^{\beta ,\alpha } C = 0$.
\end{definition}

\begin{corollary}
\label{cor3}
If $L$ and $\varphi$ do not depend on $q$,
then it follows from the fractional Hamiltonian system
(\textrm{cf.} Theorem~\ref{th3}) that
$D_{ - \gamma ;\tau }^{\beta ,\alpha } p^{\alpha,\beta}  \left( \tau  \right) = 0$,
\textrm{i.e.}, $p^{\alpha,\beta}$ is a fractional constant of motion.
\end{corollary}

\begin{definition}[fractional momentum of order $\left( {\alpha ,\beta } \right)$]
Associated with an $\left(\alpha,\beta\right)$ fractional action-like variational problem
\eqref{eq:7} we define the \emph{fractional momentum} by \eqref{eq:19}-\eqref{eq:20}:
\begin{equation*}
p^{\alpha,\beta}\left(\tau\right)
= - \frac{{\partial L}}{{\partial \dot q}}\left( {D_\gamma ^{\alpha,
\beta } q\left( \tau  \right),q\left( \tau  \right),
\tau } \right)\left( {t - \tau } \right)^{\alpha  - 1} \, .
\end{equation*}
\end{definition}

\begin{corollary}
For the fractional problem of the calculus of variations
\eqref{eq:7} with $\frac{\partial L}{\partial q} = 0$, the
fractional momentum of order $\left( {\alpha ,\beta } \right)$ is
a fractional constant of motion.
\end{corollary}


\section{Conclusions}

We generalize previous results of the fractional variational
calculus by using the action-like variational approach together
with Riemann-Liouville fractional derivatives of order
$\left(\alpha,\beta\right)$. We claim that the
$\left(\alpha,\beta\right)$ fractional action-like variational
problem here introduced offers a better mathematical model for
weak dissipative and nonconservative dynamical systems. Both
Lagrangian and Hamiltonian approaches are considered. We derive
the $\left(\alpha,\beta\right)$ fractional Euler-Lagrange,
Hamilton and Poisson equations. Standard Noetherian constants of
motion are violated due to the presence of a fractional decaying
friction term. To solve the problem we introduce a new notion of
fractional constant of motion. The class of fractional Hamiltonian
systems thus obtained has two parameters and is wider than the
standard class of Hamiltonian dynamical systems. Our approach is
different from the ones found in the literature: different from
S.~Muslih and D.~Baleanu approach \cite{[47]}, where no fractional
integral action is used and no decaying friction term considered;
different from the recent approach \cite{[48]} of A.~Stanislavsky,
where the generalization of the classical mechanics with
fractional derivatives is based on the time-clock randomization of
momenta and the coordinates are taken from the conventional phase
space.



\end{document}